\documentclass[%
 reprint,showpacs,showkeys,
 amsmath,amssymb,
 aps,
]{revtex4-1}

\usepackage{graphicx}
\usepackage{epstopdf}
\usepackage{dcolumn}
\usepackage{bm}
\usepackage{hyperref} 

\begin{document}

\preprint{APS/123-QED}

\title{Chaotic Dynamics of One-Dimensional Systems with Periodic Boundary Conditions}

\author{Pankaj Kumar}
\author{Bruce N. Miller}%
 \email{b.miller@tcu.edu}
 
\affiliation{
 Department of Physics and Astronomy,\\
 Texas Christian University, Fort Worth, TX 76129.
}

\date{\today}

\begin{abstract}
We provide appropriate tools for the analysis of dynamics and chaos for one-dimensional systems with periodic boundary conditions. Our approach allows for the investigation of the dependence of the largest Lyapunov exponent on various initial conditions of the system. The method employs an effective approach for defining the phase-space distance appropriate for systems with periodic boundary and allows for an unambiguous test-orbit rescaling in the phase space required to calculate the Lyapunov exponents. We elucidate our technique by applying it to investigate the chaotic dynamics of a one-dimensional plasma with periodic boundary. Exact analytic expressions are derived for the electric field and potential using Ewald sums thereby making it possible to follow the time-evolution of the plasma in simulation without any special treatment of the boundary. By employing a set of event-driven algorithms, we calculate the largest Lyapunov exponent, the radial distribution function and the pressure by following the evolution of the system in phase space without resorting to numerical manipulation of the equations of motion. Simulation results are presented and analyzed for the one-dimensional plasma with a view to examining the dynamical and chaotic behavior exhibited by small and large versions of the system.
\end{abstract}

\maketitle


\section{\label{sec:level1}Introduction\protect\\}

One-dimensional models serve as an effective starting point in gathering insight into more complicated higher dimensional systems. Although lower-dimensional modeling of three-dimensional systems may invoke simplifications and modifications in the mechanics and thermodynamics, one-dimensional systems have been intimately studied as a useful testing ground for approximations developed to treat the three-dimensional case. In addition, physicists have been equally interested in the intrinsic analysis of the one-dimensional systems. The idea of treating systems in one dimension is ubiquitous in the various areas of physics and a number of studies have been conducted that successfully model different phenomena in such fields as astrophysics, cosmology and plasma physics.  Consistent with the actual observations, cosmological versions of one-dimensional systems have been shown to exhibit such phenomena as hierarchical clustering and fractal behavior \cite{Rouet1990,Rouet1991} in galaxies and evaporation in black holes \cite{Russo1992}.

In plasma physics, one-dimensional models are of particular interest because they exhibit long-range forces just like the three-dimensional case. While the interactions are usually impossible to be expressed analytically in three dimensions, one dimensional modeling allows one to treat the plasma exactly while still providing profound insights. In the one-dimensional setting of plasma models where the system is comprised of parallel sheets of electric charge, the electric potential due to each sheet is given by Poisson's equation. A crucial aspect of studying such systems is to adopt appropriate boundary conditions. One of the earliest models of one-dimensional plasma studied and computer-simulated by Dawson used fixed (non-repeating) boundary conditions in which the thermalizing properties and ergodic behavior of the system were analyzed \cite{Dawson1962}. Lenard and Prager analytically studied a one-dimensional Coulomb system with non-periodic boundary conditions in the thermodynamic limit and worked out the exact statistical mechanics for systems with charges of both signs \cite{Lenard1961,Prager,EdwardLenard1962}. The two component one-dimensional plasma with an equal number of positive and negative charges has also been studied in $N$-body simulations \cite{Rouet1994,Rouet2005}. Notwithstanding the fact that these choices of boundary conditions make the mathematical analysis of thermodynamics less cumbersome, periodic boundary conditions are more appropriate in such systems because no special treatment of the boundary is required and therefore allows for the model to provide a more valid representation of a section within an extended system \cite{Springiel2006,Bertschinger1998,hockney1988,Hernquist1991}. The partition function for a special case of one-dimensional plasma with periodic boundary was worked out by Schotte and Truong for systems in which the only allowed configurations were those with charges alternating sign \cite{Schotte1980}. However, in general, an exact analytical expression for the configuration part of the partition function for systems with periodic boundary continues to be a mathematical challenge. We discuss this issue with regard to a single-component, one-dimensional periodic plasma in section \ref{thermodynamics}.

One of the most prominent reasons to understand the thermodynamics is to deduce whether or not a given system undergoes a phase transition. Although one-dimensional systems with only short-range interactions undergo no phase transition \cite{Landau1980,Mattis1993}, the possibility of a phase transition arises for versions of the system with long or infinite range interactions \cite{Hill1967_1,Hill1967_2,Dyson1969,Pirjol2014}. For the particular case of a two-component one-dimensional periodic plasma with alternating charge configuration, Schotte showed analytically that the system underwent a second order phase transition \cite{Schotte1980}.  However, due to the lack of exact analytical statistical mechanics for a general case of periodic plasma or gravitational systems, one has to look for other reliable indications of phenomena that are normally deduced by formulating exact statistical mechanics.

Lyapunov exponents have been shown to act as indicators of phase transitions \cite{Butera1987,Caiani1997,Casetti2000}. For systems with well-behaved trajectories, Lyapunov exponents can be determined by studying the geometry of the phase-space trajectories \cite{Caiani1997,Casetti2000}. However, if the trajectories are not well-behaved, the tangent-space is not defined at all points on the manifold and the calculation of the Lyapunov exponents calls for numerical investigation of the divergence of the phase-space trajectories in response to small perturbations. The numerical simulation of Lyapunov exponents has been done for the one-dimensional case and it has been shown that a system with a finite number of particles can exhibit properties that are observed in the thermodynamic limit \cite{Bonasera1995}. In the case of a two-dimensional particle system, it has been shown that the largest Lyapunov exponent exhibits a maximum at the fluid-solid phase transition \cite{Dellago1996}. Moreover, for a one-dimensional chain of coupled nonlinear oscillators, the largest Lyapunov exponent has also been shown to exhibit phase-transition-like behavior similar to the one exhibited by the system itself near the transition temperature \cite{Barré2001}.

Apart from being a possible indicator of phase transitions, the largest Lyapunov exponent is itself of great interest to physicists; in the study dynamical systems, chaotic instability can be unambiguously quantified using the Lyapunov exponents \cite{Benettin1980, ott2002}. Of particular interest is the value of the largest (maximal) Lyapunov exponent which is usually sufficient to quantify the degree of chaos \cite{sprott2003}, especially in non-dissipative systems. For non-periodic systems, the largest Lyapunov exponent can be calculated numerically provided that one has the exact time evolution of the positions and velocities of the particles. However, for a system with periodic boundary, care must be taken in defining the phase-space distance and the vector in the direction of phase-space separation between the reference and the test orbits. We deal with the issue of phase-space rescaling for the one-dimensional periodic systems in Sec. \ref{phasespace}.

In order to follow the dynamics of the particles and study the system using a molecular dynamics approach, one first needs an efficient algorithm free from any special treatment of the boundary. Such a technique was devised by Miller and Rouet for the case of a one-dimensional periodic gravitational system based on their derivations of the gravitational field and potential \cite{miller2010}. In the present work, we extend the approach by deriving the electric field and potential for the case of a plasma and use the resulting equations of motion to follow the motion of the particles and calculate the largest Lyapunov exponents (Sec. \ref{1DPlasma} and \ref{chaos}). To elucidate the approach, we carefully investigate the dynamics and chaos of a system of four particles in simulation (Sec. \ref{simulation_4P}). We also demonstrate the utility of our approach in understanding the chaotic dynamics of larger one-dimensional systems with periodic boundary by treating a plasma system with forty particles (Sec. \ref{simulation_40P}). Finally, in Sec. \ref{discussions}, we discuss and compare the simulation results obtained for the case of the four-particle and the forty-particle systems with a view to examining the behavior of the plasma for small and large number of particles. 

\section{The one-dimensional plasma} \label{1DPlasma}

We consider a periodic system of $N$ discrete sheets with surface mass and charge densities $m$ and $q$ respectively with the primitive cell located in $-L\leq x\leq L$. The cell is immersed in a uniformly charged background, $\rho_0={-Nq}/{2L}$; the net charge in the cell is zero. Hence the charge density as a function of the position in the cell can be given as
\begin{equation} \label{eq_charge}
\rho_p(x) = q \sum\limits_{j=1}^N \left[\delta(x-x_j) - \frac{1}{2L}\right] ,
\end{equation}
where $x_j$ is the position of the $j$-th particle in the primitive cell. The above representation of the charge distribution suggests that every discrete sheet (henceforth referred to as a particle) in the primitive cell carries a neutralizing charge density of magnitude $q/2L$ distributed uniformly in the cell.

\subsection{Electric Potential and Field}
Analogous to the gravitational case treated by Miller and Rouet \cite{miller2010}, the charge density in the plasma system repeats periodically and extends in the entire one-dimensional space. Normally, the electric potential can be found by integrating the Poisson equation. In the case of plasma, this would entail integrating the charge density in Eq. (\ref{eq_charge}). One could attempt to calculate the integrand by treating separately the contributions from the individual particles and then adding the contribution from the uniform background. However, such a treatment produces divergent integrands for the case of an infinite system. To circumvent this difficulty, we follow the approach proposed by Kiessling \cite{Kiessling2003}. We separate the charge contributions from the discrete sheets and the background and treat the two resulting integrands by applying a screening function. Following this approach, the screened potential contribution from the neutralizing background alone due to a particle located at $x_1$ and its replicas in the periodic system is found as:
\begin{equation} \label{eq_kiess_background}
\Psi_{b_{1}}(x,\kappa) = \frac{-\pi kq}{L}\int_{-\infty}^{+\infty}|x-x'| e^{\left(-\kappa|x-x'|\right)}  dx',
\end{equation}
where $k = \frac{1}{4\pi \epsilon_o}$ and $\kappa > 0$ is a small screening parameter. Mathematical simplification of Eq. (\ref{eq_kiess_background}) yields
\begin{equation} \label{eq_kiess_background_sum}
\Psi_{b_{1}}(x,\kappa) = \frac{2\pi kq}{L\kappa^2}.
\end{equation}
The screened potential contribution due to the particle at $x_1$ and its replicas can then be found as
\begin{equation} \label{eq_kiess_sheet}
\Psi_{\sigma_{1}}(x,\kappa) = {2\pi kq}\int_{-\infty}^{+\infty}S_{{\delta}_1}(x) |x-x'| e^{\left(-\kappa|x-x'|\right)}  dx'
\end{equation}
where
\begin{equation} \label{eq_S_delta}
S_{{\delta}_1}(x) = \sum\limits_{r=-\infty}^{\infty} \delta(x-(x_1+2rL)) .
\end{equation}
Eq. (\ref{eq_kiess_sheet}) reduces to
\begin{equation} \label{eq_kiess_sheet_sum}
\Psi_{\sigma_{1}}(x,\kappa) = {-2\pi kq}\left\lbrace{S_{{<}_1}(x,\kappa) + S_{{>}_1}(x,\kappa)}\right\rbrace
\end{equation}
where
\begin{equation} \label{S1}
S_{{<}_1}(x,\kappa) = \sum\limits_{r=-\infty}^0 \left\lbrace{x-\left(x_1 + 2rL\right)}\right\rbrace e^{\left\lbrace-\kappa(x-x_1-2rL)\right\rbrace}
\end{equation}
and
\begin{equation} \label{S2}
S_{{>}_1}(x,\kappa) = \sum\limits_{r=1}^{\infty} \left\lbrace{\left({x_1 + 2rL}\right)-x}\right\rbrace e^{\left\lbrace-\kappa(x_1+2rL-x)\right\rbrace}.
\end{equation}
Adding Eqs. (\ref{eq_kiess_background_sum}) and (\ref{eq_kiess_sheet_sum}), we obtain the net screened potential due to the particle at $x_1$ (including its replicas and the associated neutralizing background) as
\begin{equation} \label{eq_net_screened_potential}
\Psi_{1}(x,\kappa) = {-2\pi kq}\left[S_{{<}_1}(x,\kappa) + S_{{>}_1}(x,\kappa) - \frac{1}{L\kappa^2}\right]
\end{equation}
Finally, the electric potential at a position $x$ due to a particle at $x_1$ in the primitive cell (and its replicas) can be found by evaluating $\Psi_{1}(x,\kappa)$ in the limit $\kappa \rightarrow 0$ as
\begin{equation} \label{eq_potential1}
\Phi_1(x) = 2\pi kq \left[\frac{(x-x_1)^2}{2L} - |x-x_1| - \frac{L}{3}\right]
\end{equation}
Hence the field at the position $x$ because of the particle at $x_1$ and its replicas can be determined as
\begin{equation} \label{eq_field1}
E_1(x) = \left\lbrace
\begin{array}{l r}
\left.2\pi kq \left(1 - \frac{(x-x_1)}{L}\right)\right., & x>x_1\\
\;\\
2\pi kq \left(-1 - \frac{(x-x_1)}{L}\right), & x<x_1
\end{array}\right.\;
\end{equation}
In order to find the net field at a point $x$ due to the entire system, we sum up the field contribution from all the $N$ particles. We get
\begin{equation} \label{eq_netField}
E(x) = 2\pi kq \left\lbrace \frac{N}{L}(x_c-x) + N_{left}(x) - N_{right}(x)\right\rbrace,
\end{equation}
where $x_c$ is the position of the center of mass of the primitive cell with $N_{left}(x)$ and $N_{right}(x)$ representing respectively the numbers of particles to the left and to the right of position $x$ in the primitive cell.
\subsection{Modes of Oscillations and Thermodynamics} \label{thermodynamics}
The Hamiltonian in terms of the potential energy, $V$ and  momenta, $p_j$ is given by
\begin{equation} \label{eq_Hamiltonian}
{\cal H} = V + \frac{1}{2m}\sum\limits_{j=1}^N p_j^2
\end{equation}
where
\begin{equation} \label{eq_potentialEnergy}
V = \frac{1}{2}\sum\limits_{i=1}^N \sum\limits_{j=1}^N q \Phi_i(x_j) .
\end{equation}\\
Without the loss of generality, if we assume that the center of mass of the primitive cell is located $x=0$, then from Eq. (\ref{eq_netField}), we obtain $N$ equilibrium positions given by 
\begin{equation} \label{eq_eqmPos}
X_j=\left(2j-N-1\right)(L/N) .
\end{equation} 
We then express each of the $x_j$ as its displacement relative to the corresponding equilibrium position $X_j$ i.e., $x_j = X_j + y_j$. Under this transformation of variables, the potential energy of the system takes the form
\begin{equation} \label{eq_potentialNorm}
V = V_o + \frac{2\pi kq^2}{L}\left\lbrace\left(\frac{N-1}{2}\sum\limits_{j=1}^N y_j^2\right)-\frac{1}{2}\sum\limits_{j\neq i=1}^N y_i y_j \right\rbrace
\end{equation}
where $V_o$ is a constant which has no bearing on the equations of motion. Eq. (\ref{eq_potentialNorm}) can also be expressed as
\begin{equation} \label{eq_pot_mat_eqn}
V = V_o + Y^{\dagger}  \hat{V}  Y,
\end{equation}
where $Y^{\dagger} = (y_1,y_2,\hdots,y_N)$ and $\hat{V}$ is the coupling matrix that takes the $(N \times N)$ Toeplitz form
\begin{equation} \label{eq_potentialMat}
\hat{V} = \frac{\pi kq^2}{L}
\left(
\begin{matrix}
N-1 &  -1  &  -1  & \ldots & -1\\
-1 &  N-1  &  -1  & \ldots & -1\\
-1 &  -1  &  N-1  & \ldots & -1\\
\vdots & \vdots & \vdots & \ddots & \vdots\\
-1 & -1  & -1 & \ldots & N-1
\end{matrix}
\right) .
\end{equation}
The matrix corresponding to the kinetic energy is already diagonal and is expressed simply as $\hat{T}=mI_n$ where $I_n$ is the $N$-dimensional identity matrix. The solution to the  eigenvalue problem yields only two unique eigenvalues of the Hamiltonian: $\omega_o^2=0$, which is non-degenerate with eigenvector whose elements are identical, and, $\omega_1^2={\frac{2\pi kq^2N}{mL}}$ which is $(N{-}1)$-fold degenerate with eigenvectors lying in the subspace formed by the columns of the $N\times (N-1)$ matrix 
\begin{equation} \label{eq_Omega}
\Omega =
{\left(
\begin{matrix}
-1 &  -1  &  -1  & \ldots & -1\\
1 &  0  &  0  & \ldots & 0\\
0 &  1  &  0  &   \ldots & 0\\
0 &  0  &  1  &   \ldots & 0\\
\vdots & \vdots & \vdots & \ddots & \vdots\\
0 & 0  &  0  &  \ldots & 1
\end{matrix}
\right)}_{N\times (N-1)} .
\end{equation}
The eigenvector corresponding to the non-degenerate eigenvalue simply conforms to the situation in which every particle is equidistant from its corresponding equilibrium position and has the same velocity as other particles. If this is the sole mode excited, there will be no oscillation as the particles remain in mechanical equilibrium at all times. This mode is basically a consequence of the conservation of linear momentum for the system.

The canonical partition function for the system is given by
\begin{eqnarray} \label{eq_PartitionFunction}
Z_N = \frac{1}{h^N}\int_{-\infty}^{+\infty} &&\hdots \int dp_1 \hdots dp_N \int_{-L}^{+L} \hdots \hdots \nonumber \\
&&\hdots \int dx_1 \hdots dx_N \hspace{2 pt} exp(-{\cal{H}}/kT).
\end{eqnarray}
In several non-periodic cases \cite{Lenard1961,Prager,EdwardLenard1962,Baxter1963,Ruelle1968} the  partition function can be dealt with analytically in the thermodynamic limit. In particular, Baxter's model of a one-dimensional, single component, non-periodic plasma with uniform background allowed for integration in Eq. (\ref{eq_PartitionFunction}) by performing a change of coordinates to normal modes which corresponded to a set of coordinates whose integration limits could be found easily for an ordered system. A combinatorial factor was included in the result to account for various possible permutations of the particles \cite{Baxter1963}. However, in our case, the normal modes, in general, are a linear combination of the coordinates of all the particles. The most obvious choice is to look for normal coordinates that follow some kind of cyclic pattern or permutation with respect to one another. For versions of our system with even number of particles, it is possible to find the $(N-1)$ degenerate normal modes that are linear combinations of individual particle coordinates in which half of the coordinates appear as their additive inverses. For example, if we look for normal modes for a four-particle system, the three degenerate eigenvectors will lie in the subspace formed by Eq. (\ref{eq_Omega}) for which one of the possible sets of orthonormal coordinates is given by 
\begin{equation}
Q_1 = \frac{1}{2}\left(
\begin{matrix}
1\\
1\\
1\\
1
\end{matrix} \right)
\end{equation}
\begin{equation}
Q_2 = \frac{1}{2}\left(
\begin{matrix}
1\\
-1\\
1\\
-1
\end{matrix} \right), \hspace{4 pt}
Q_3 = \frac{1}{2}\left(
\begin{matrix}
1\\
1\\
-1\\
-1
\end{matrix} \right), \hspace{4 pt}
Q_4= \frac{1}{2}\left(
\begin{matrix}
1\\
-1\\
-1\\
1
\end{matrix} \right)
\end{equation}
One can easily determine that there is no simple way to find the limits of integration for these normal coordinates, even in the case of a configuration of ordered particles. It is this difficulty that makes the analytical treatment of the thermodynamics of the periodic versions of one-dimensional plasmas and gravitational systems an area of research that has been relatively unexplored.
\subsection{Equations of Motion and Dynamics} \label{EqnMotion}
The equation of motion for the $j$-th particle located at position $x_j$ under the action of the net field given by Eq. (\ref{eq_netField}) is
\begin{equation}
\frac{m}{q} \frac{d}{dt} (v_j) = 2\pi kq \left\lbrace \frac{N}{L}(x_c-x) + N_{left}(x) - N_{right}(x)\right\rbrace
\end{equation}
The motion of the $(j+1)$-th particle relative to that of the $j$-th particle can then be expressed as
\begin{equation}
\frac{m}{q} \frac{d}{dt} (v_{j+1}-v_j) = -2\pi kq \left\lbrace \frac{N}{L}(x_{j+1}-x_{j}) -2\right\rbrace
\end{equation}\\
Defining $Z_j \equiv (x_{j+1}-x_j)$ and $W_j \equiv (v_{j+1}-v_j)$,
\begin{equation} \label{eq_relMotion}
\frac{m}{q} \frac{d}{dt} (W_j) = -2\pi kq \left\lbrace \frac{N}{L}Z_j -2\right\rbrace
\end{equation}
Solutions to Eq. (\ref{eq_relMotion}) provide the displacement and velocities of $(j+1)$-th particle with  respect to $j$-th particle and are expressed simply as a function of time:
\begin{equation}
Z_j (t)=\frac{2L}{N}+\left\lbrace Z_j (0)-\frac{2L}{N}\right\rbrace \cos{\omega}t + \frac{Z_j'(0)}{{\omega}}\sin{\omega}t
\end{equation}
\begin{equation}
W_j (t)=-{\omega}\left\lbrace Z_j (0)-\frac{2L}{N}\right\rbrace \sin {\omega}t + Z_j'(0)\cos{\omega}t
\end{equation}
where $\omega \equiv \sqrt{\frac{2\pi kq^2N}{mL}}$ is the plasma frequency. Crossing, if any, between $j$-th and $(j+1)$-th particle will correspond to $Z_j(t) = 0$. Actual positions can be obtained from solving the set of simultaneous equations given by
\begin{equation}
\left(
\begin{matrix}
-1 &  1  &  0  &  0  &   \ldots & 0 & 0\\
0 &  -1  &  1  &  0  &   \ldots & 0 & 0\\
0 &  0  &  -1  &  1  &   \ldots & 0 & 0\\
\vdots & \vdots & \vdots  & \vdots & \ddots & \vdots & \vdots\\
0 &  0  &  0  &  0    & \ldots & -1 & 1\\
1 &  1  &  1  &  1   & \ldots & 1 & 1
\end{matrix}
\right)
\left(
\begin{matrix}
x_1\\
x_2\\
x_3\\
\vdots\\
x_{N-1}\\
x_{N}
\end{matrix}
\right)
=
\left(
\begin{matrix}
Z_1\\
Z_2\\
Z_3\\
\vdots\\
Z_{N}\\
Nx_c
\end{matrix}
\right)
\end{equation}
Velocities can be obtained in a similar way by solving the simultaneous set of equations involving $W_j$s. It must be noted that the only unique non-zero value of the frequency obtained from the normal mode analysis is the same as that obtained from the equation of motion via summing the field contributions from individual particles. Interestingly, since each of the $(N-1)$  values of $Z_j$ can be  expressed as a linear combination of the column vectors constituting $\Omega$ in Eq. (\ref{eq_Omega}), every $Z_j$ is an eigenvector of the Hamiltonian with eigenvalue, $\omega_1^2$. Hence the two methods, namely, the sum over individual field contributions and the normal-mode approach, produce consistent results. The former is more effective in dealing with a system that undergoes crossings since the method allows for iterative redefinition of the system after every crossing. The normal mode method, on the other hand, allows one to express the total Hamiltonian (of a system with no crossings) as a sum of those of $N$ decoupled oscillators. Decoupling can easily be performed by finding an orthonormal basis in the $(N - 1)$-dimensional subspace represented by the columns of $\Omega$ in Eq. (\ref{eq_Omega}).
\section{Stability and Chaos} \label{chaos}
\subsection{The Largest Lyapunov Exponent}
The degree of chaos in a system can be quantified by finding the largest Lyapunov exponent for the system.  Two trajectories in phase space with a small separation will diverge for chaotic dynamical systems and the rate of separation of the trajectory is characterized by the largest Lyapunov exponent. We start with a reference system with given initial position in the phase space. A test system is then defined by perturbing the reference system. As the two systems are allowed to evolve in time, the distance between their trajectories will normally change rapidly without any bounds for chaotic systems which might result in a computer overflow \cite{ott2002}. This problem is overcome by rescaling the phase-space distance between the reference and the test trajectories periodically to the initial separation. The rescaling is done in such a way that the relative direction of the test-orbit position is unchanged with respect to the position of the reference in the phase space \cite{sprott2003}. Mathematically, the largest Lyapunov exponent can be expressed as
\begin{equation} \label{eq_LyapExponent}
\lambda = \lim_{l \to \infty}\frac{1}{l\tau}\sum\limits_{i=0}^l ln\left(\frac{d_i}{d_o}\right)
\end{equation}
where $d_o$ is the initial separation, $d_i$ is the separation after $i$-th iteration, $\tau$ is the time interval between iterations, and  $l$ is the total number of iterations \cite{ott2002}.

\subsection{Phase Space for One-Dimensional Systems with Periodic Boundary} \label{phasespace}
One of the key features of the one-dimensional plasma with periodic boundary is that the particles can be thought of as being located on a torus of circumference $2L$. In order to avoid sudden discontinuities arising from boundary crossings, the interaction between any two particles at positions $x_j$ and $x_{j+1}$ in the primitive cell must correspond to the minimum of $|x_{j+1}-x_j|$ (the spatial distance in the primitive cell) and $(2L-{|x_{j+1}-x_j|})$. Hence the phase-space separation between a given reference system, $R$ and a perturbed system, $T$ will be given as
\begin{figure}[t]
\includegraphics[scale=0.6]{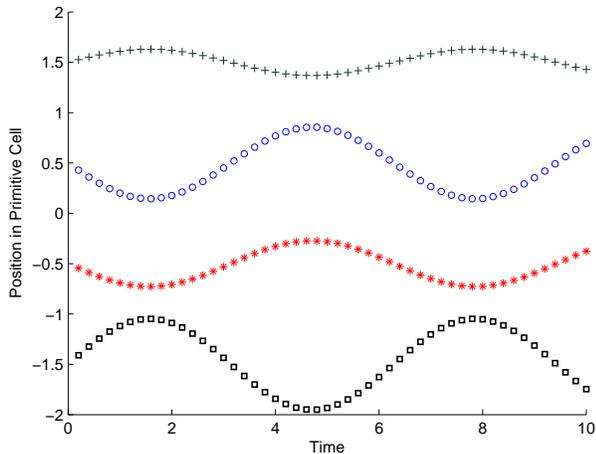}
\caption{\label{fig:Dyn4PLowEnergy} Position vs. Time for a plasma system of four particles with low energy. The particles perform pure oscillations about the equilibrium positions when the total energy is low. No interparticle crossings occurs in this low energy state and the order of the particles is maintained on the torus.}
\end{figure}
\begin{figure}[t]
\includegraphics[scale=0.6]{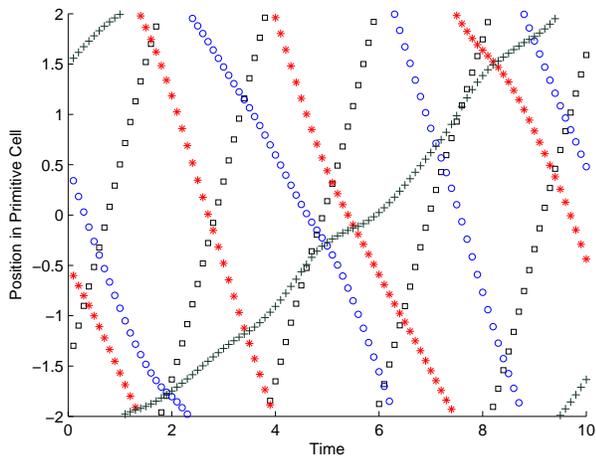}
\caption{\label{fig:Dyn4PHighEnergy} Position (in the primitive cell) vs. Time for the four particle system with high energy. The particles cross each other and the boundary. When a particle crosses one of the boundaries in the periodic system, it emerges at the other.}
\end{figure}
\begin{figure}[t]
\includegraphics[scale=0.6]{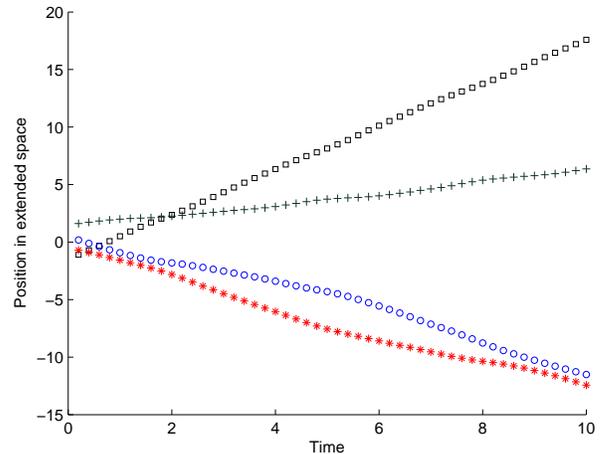}
\caption{\label{fig:Dyn4PHighEnergyExtSp} Position vs. Time when tracked in the extended space for the four-particle system shown in Fig. \ref{fig:Dyn4PHighEnergy}. At high energies, the kinetic energy is dominant and the particles feel the potential as a mere perturbation thereby making the motion tend to the one with constant velocity.}
\end{figure}
\begin{figure*}[t]
\includegraphics[scale=0.8]{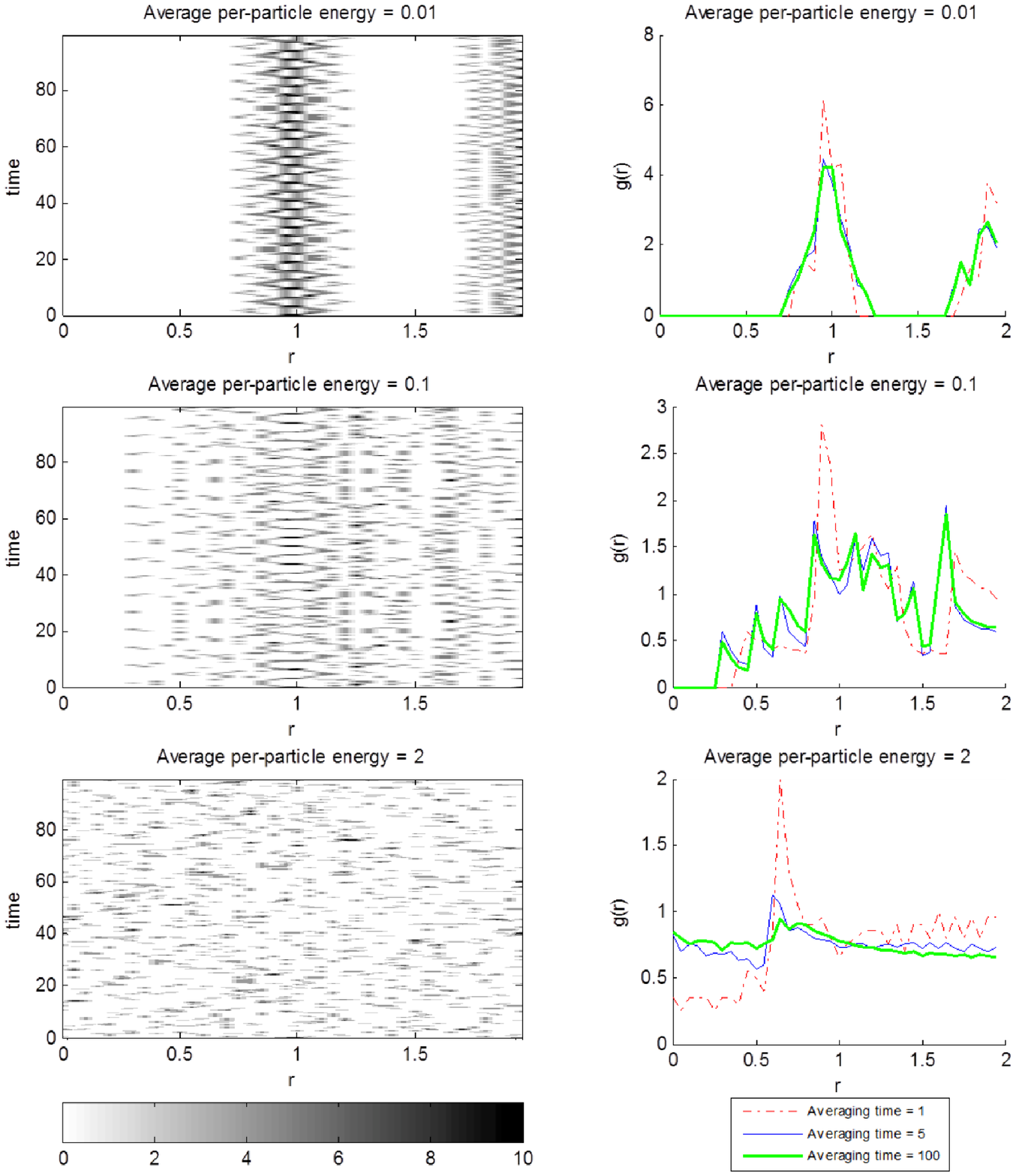}
\caption{\label{fig:CorrelationFunction} Pair correlation function for the four particle system with different per-particle energies. The figures on the left show the time evolution whereas the plots on the right are the time-averaged values for three different averaging times.}
\end{figure*}
\begin{figure}[t]
\includegraphics[scale=0.6]{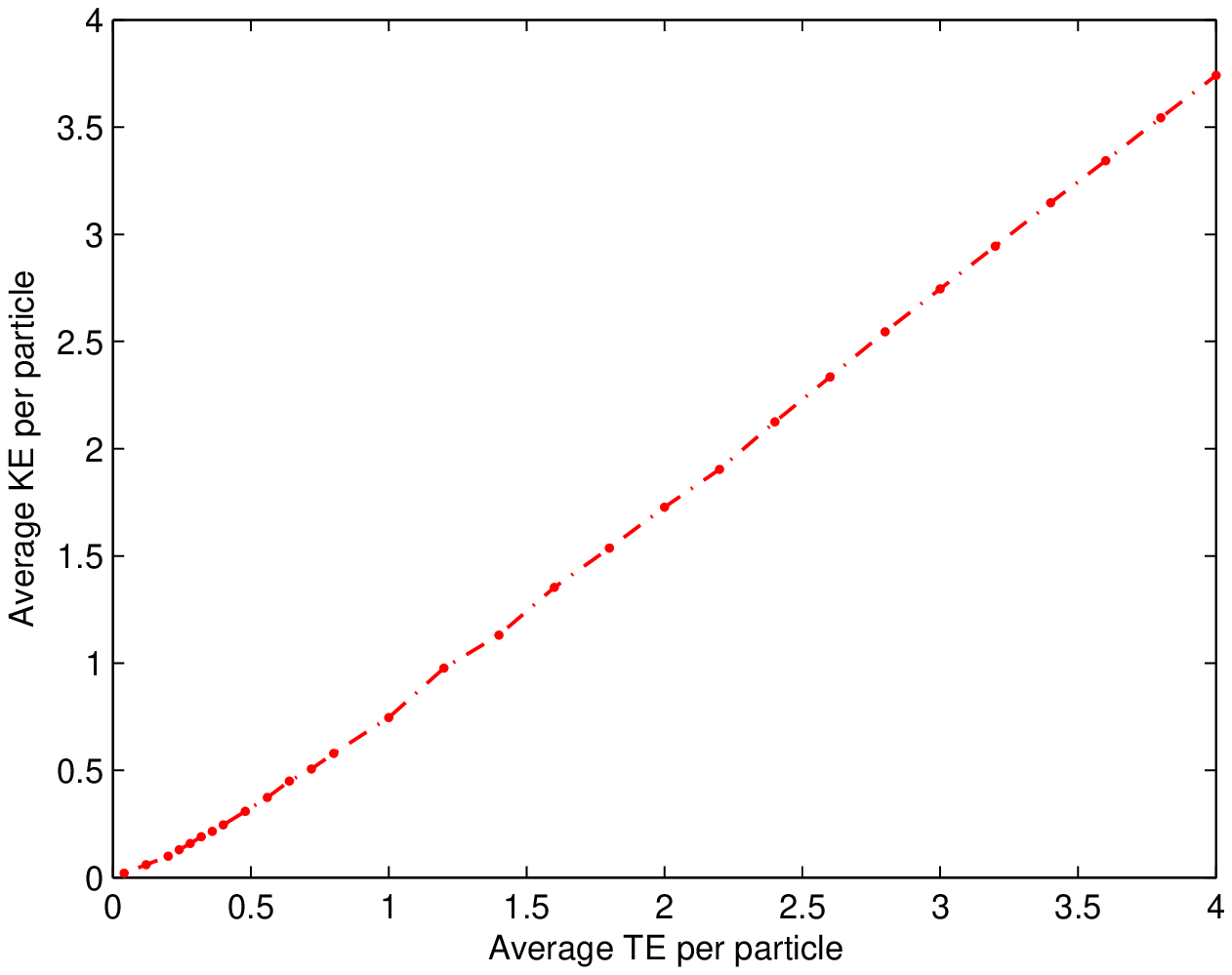}
\caption{\label{fig:KEvsTE4P} Average per-particle kinetic energy plotted against average per-particle total energy for the system of four particles.}
\end{figure}
\begin{widetext}
\begin{equation} \label{eq_dprim}
d_{prim}=\sqrt{\sum\limits_{j=1}^N\left(min\left\lbrace|x_{j_R}-x_{j_T}|,\left(2L-|x_{j_R}-x_{j_T}|\right)\right\rbrace\right)^2+ \sum\limits_{j=1}^N\left(v_{j_R}-v_{j_T}\right)^2}
\end{equation}
\end{widetext}
However, by confining the positions to the primitive cell and calculating the separation as given in Eq. (\ref{eq_dprim}) leads to ambiguity in the direction of the vector representing the relative position of the test system with respect to the reference system. An example of such a problem is the situation whereupon a particle at $x_{j_T}$ in the test system crosses the boundary whereas its counterpart in the reference system at $x_{j_R}$ remains on the original side; when a particle leaves the primitive cell of the periodic system, it emerges at the other end of the cell. Hence, restricting the representation of the system coordinates to the primitive cell results in an abrupt switch in the sign of the spatial components of the phase-space separation vector.

We approach the solution by first acknowledging the validity of the magnitude of the separation vector expressed in Eq. (\ref{eq_dprim}). Hence, the only job at hand is to define a unit vector in phase-space that preserves the relative directional information between the two trajectories under events like boundary crossings or a sudden change in particle ordering as a result of rescaling. We achieve this goal by tracking the position of every particle in the extended one-dimensional spatial coordinates (not restricted to the primitive cell). In this case, the center of the primitive cell becomes the origin of the one-dimensional space and the particles’ positions are tracked in the extended space in which their position coordinates are not restricted to within the interval $[−L, L)$. The required unit vector is found as
\begin{equation} \label{eq_unitVector}
\hat{X}_{T_{R}} = \frac{1}{d_{prim}}
\left(
\begin{matrix}
{\left({\tilde{x}}_{1_{T}} - {\tilde{x}}_{1_{R}}\right)}\\
\vdots\\
{\left({\tilde{x}}_{N_{T}} - {\tilde{x}}_{N_{R}}\right)}\\
\left(v_{1_{T}} - v_{1_{R}}\right)\\
\vdots\\
\left(v_{N_{T}} - v_{N_{R}}\right)\\
\end{matrix}
\right)
\end{equation}
where $\tilde{x}$ represent the positions in the extended one-dimensional space. Examples of time evolution of $x$ and $\tilde{x}$ have been discussed in Sec. \ref{simulation_4P} (Figs. \ref{fig:Dyn4PHighEnergy} and \ref{fig:Dyn4PHighEnergyExtSp} respectively show the positions in primitive cell ($x$) and positions in extended space, ($\tilde{x}$) as functions of time for a system of four particles with energy sufficient to allow for boundary crossings).

Once the unit vector has been found from Eq. (\ref{eq_unitVector}), the rescaled test orbit may be determined in terms of the initial separation, $d_o$ as
\begin{equation} \label{eq_testOrbit}
\vec{X}_T = \vec{X}_R + d_o \hat{X}_{T_{R}}
\end{equation}
\begin{figure}[h]
\includegraphics[scale=0.6]{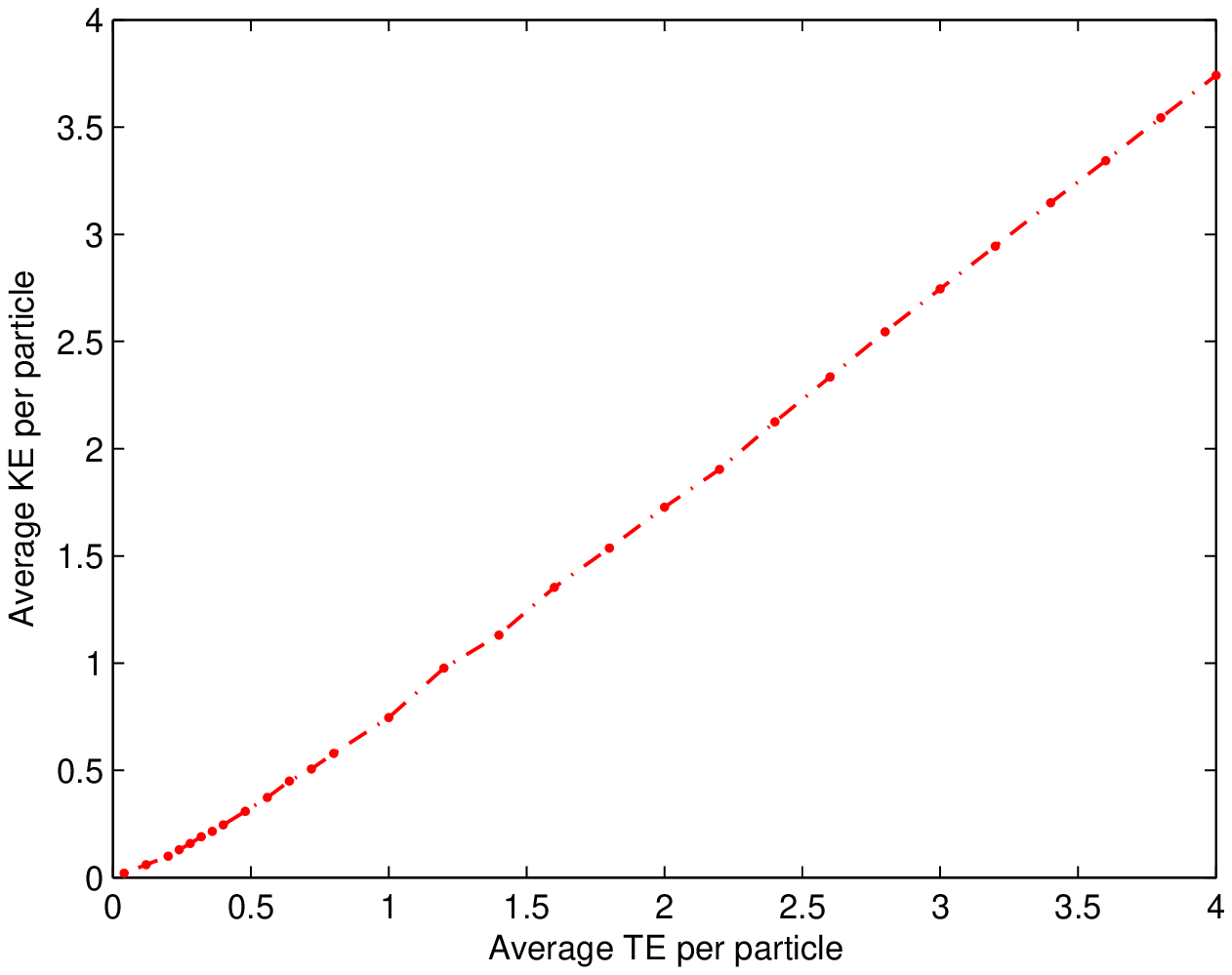}
\caption{\label{fig:PEvsTE4P} Average per-particle potential energy plotted against the average per-particle kinetic energy for the system of four particles.}
\end{figure}
\begin{figure}[h]
\includegraphics[scale=0.6]{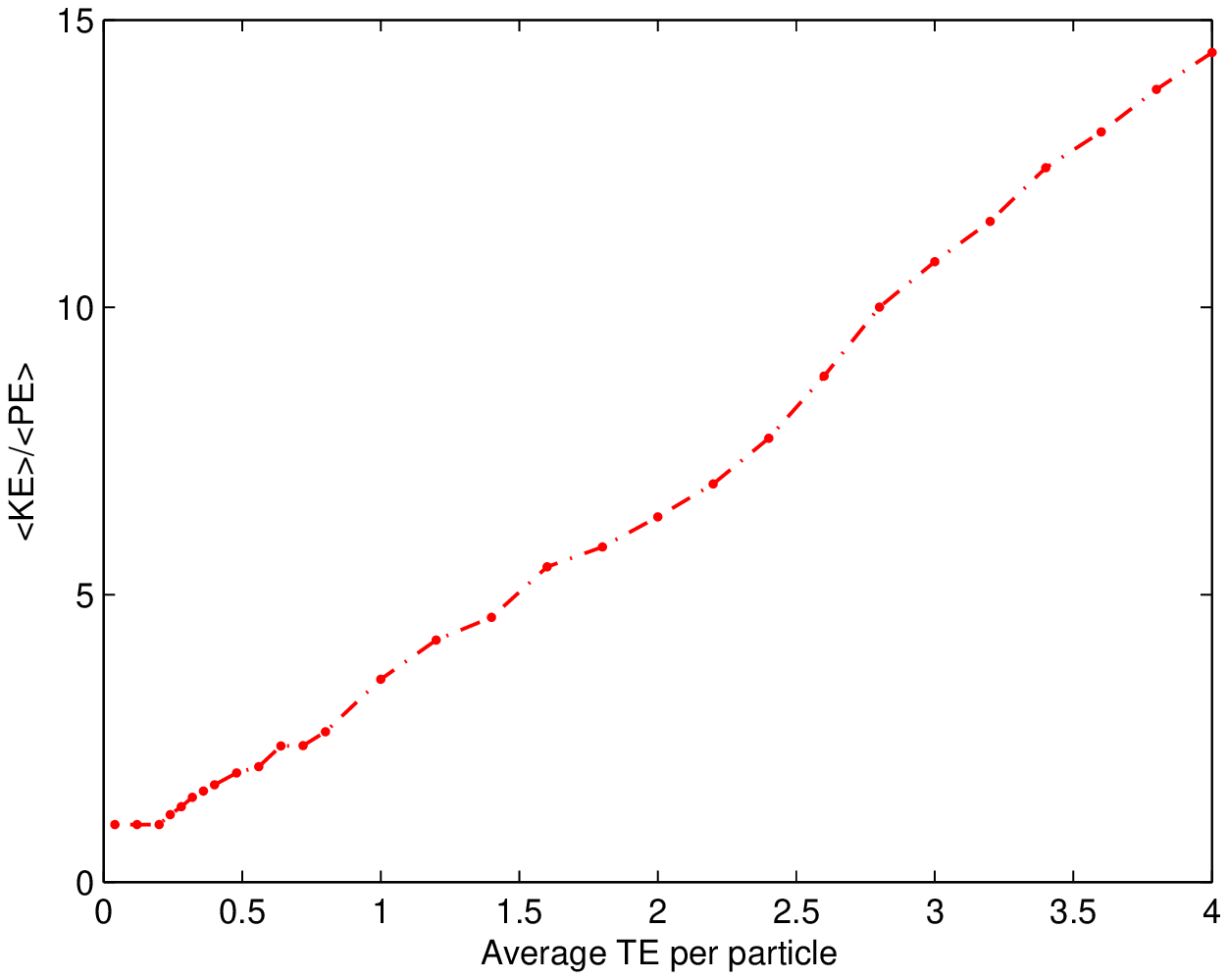}
\caption{\label{fig:VirialVsTE4P} Time averaged virial ratio plotted againt average per-particle energy for the system of four particles.}
\end{figure}
\begin{figure}[h]
\includegraphics[scale=0.6]{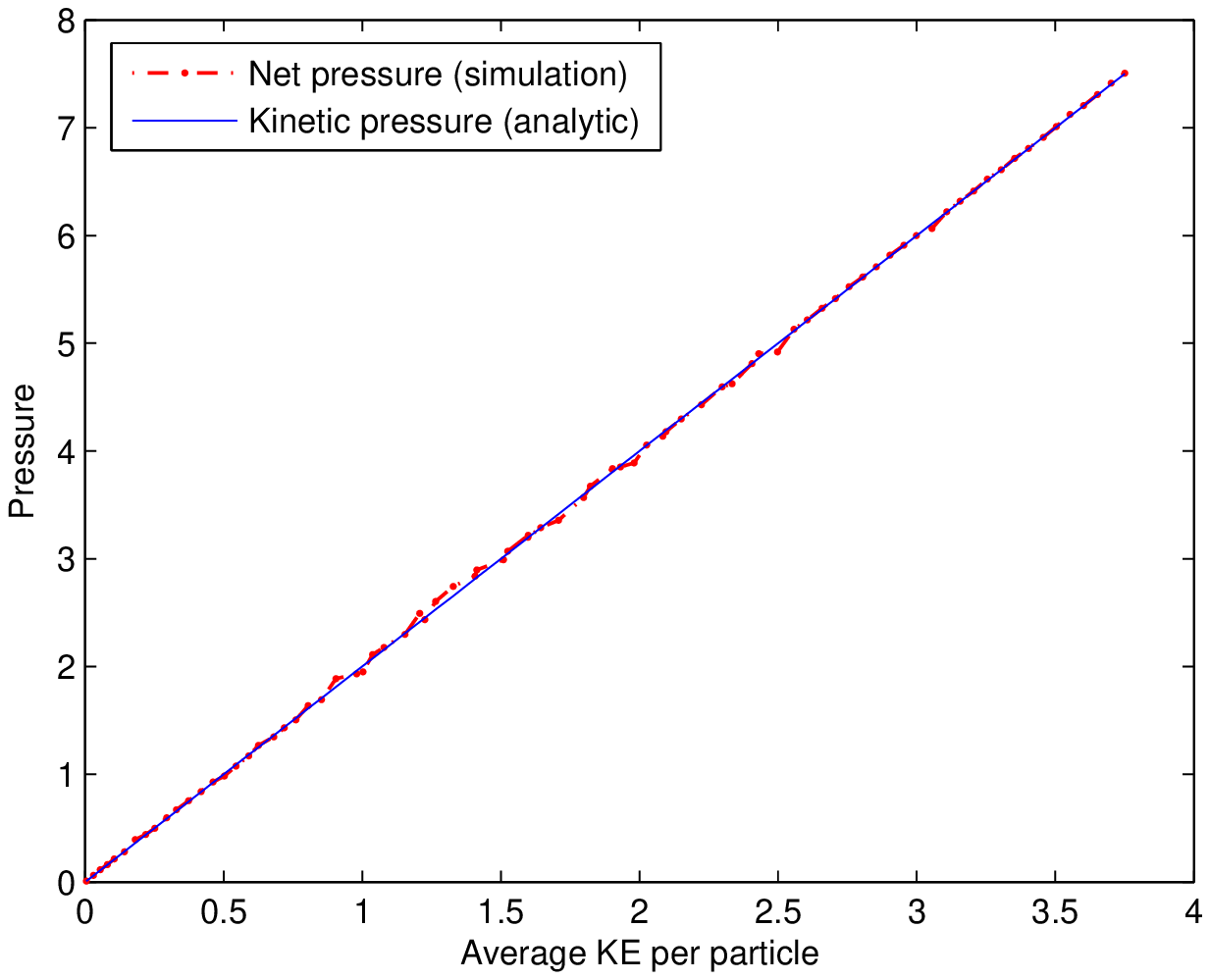}
\caption{\label{fig:PressVsEnergy4P} Average net pressure (from simulation) and kinetic pressure (analytic) plotted against average per-particle kinetic energy for the four-particle system.}
\end{figure}
\begin{figure}[h]
\includegraphics[scale=0.6]{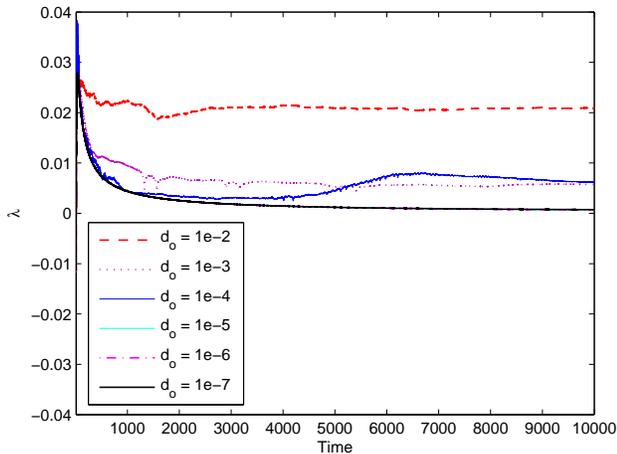}
\caption{\label{fig:LyapConv4P} An example of the convergence dependence of the largest Lyapunov exponent on the size of the test-orbit perturbation ($d_o$) for the four-particle system. The plots corresponding to the perturbation sizes 1e-5, 1e-6, and 1e-7 show close resemblance in their behavior suggesting that a decrease in the value of $d_o$ below 1e-5 does not result in any considerable change in the converged value of the largest Lyapunov exponent.}
\end{figure}
\begin{figure}[h]
\includegraphics[scale=0.6]{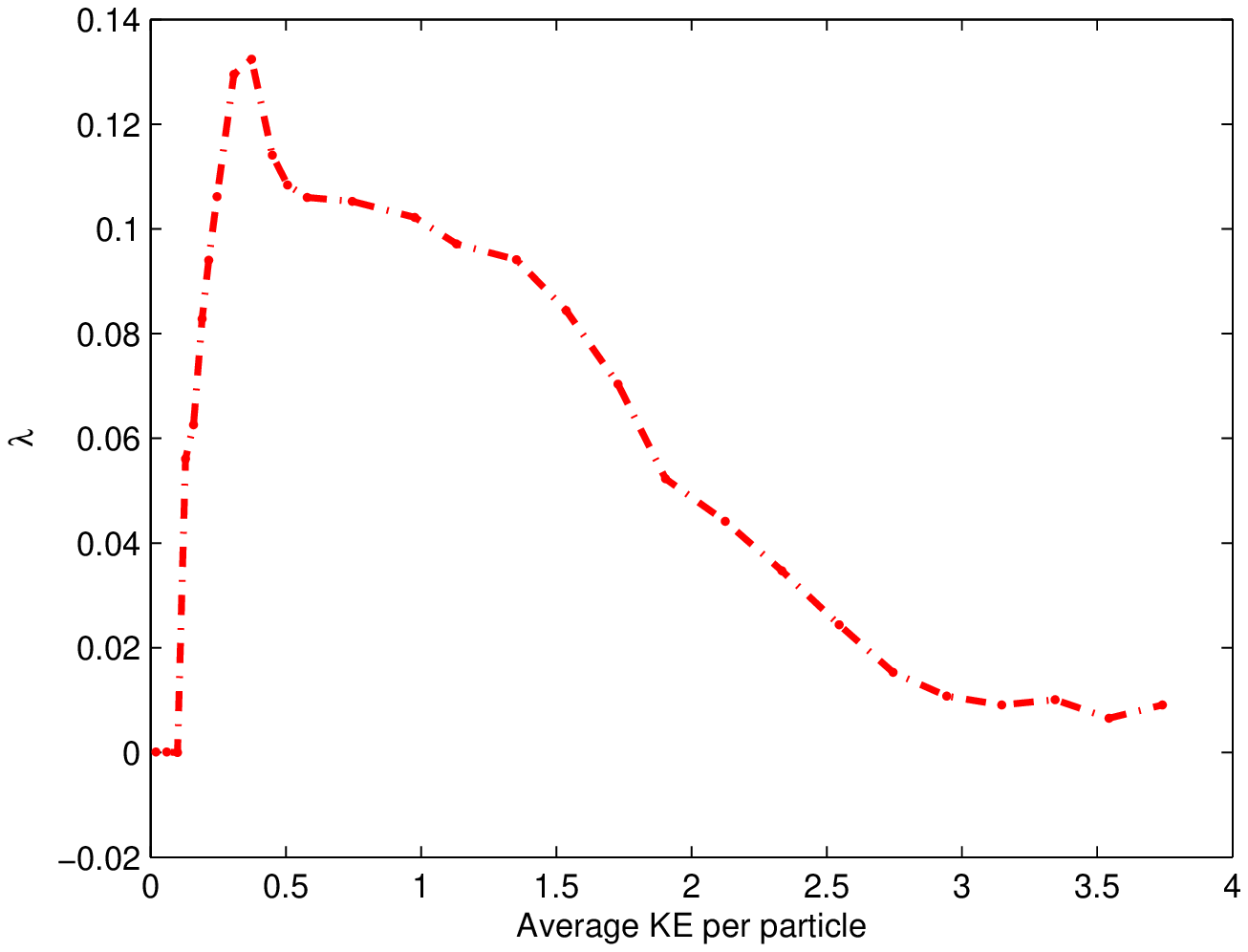}
\caption{\label{fig:LyapVsEnergy4P} Largest Lyapunov exponent plotted against average per-particle kinetic energy for a four particle system.}
\end{figure}
\begin{figure*}[t]
\includegraphics[scale=0.9]{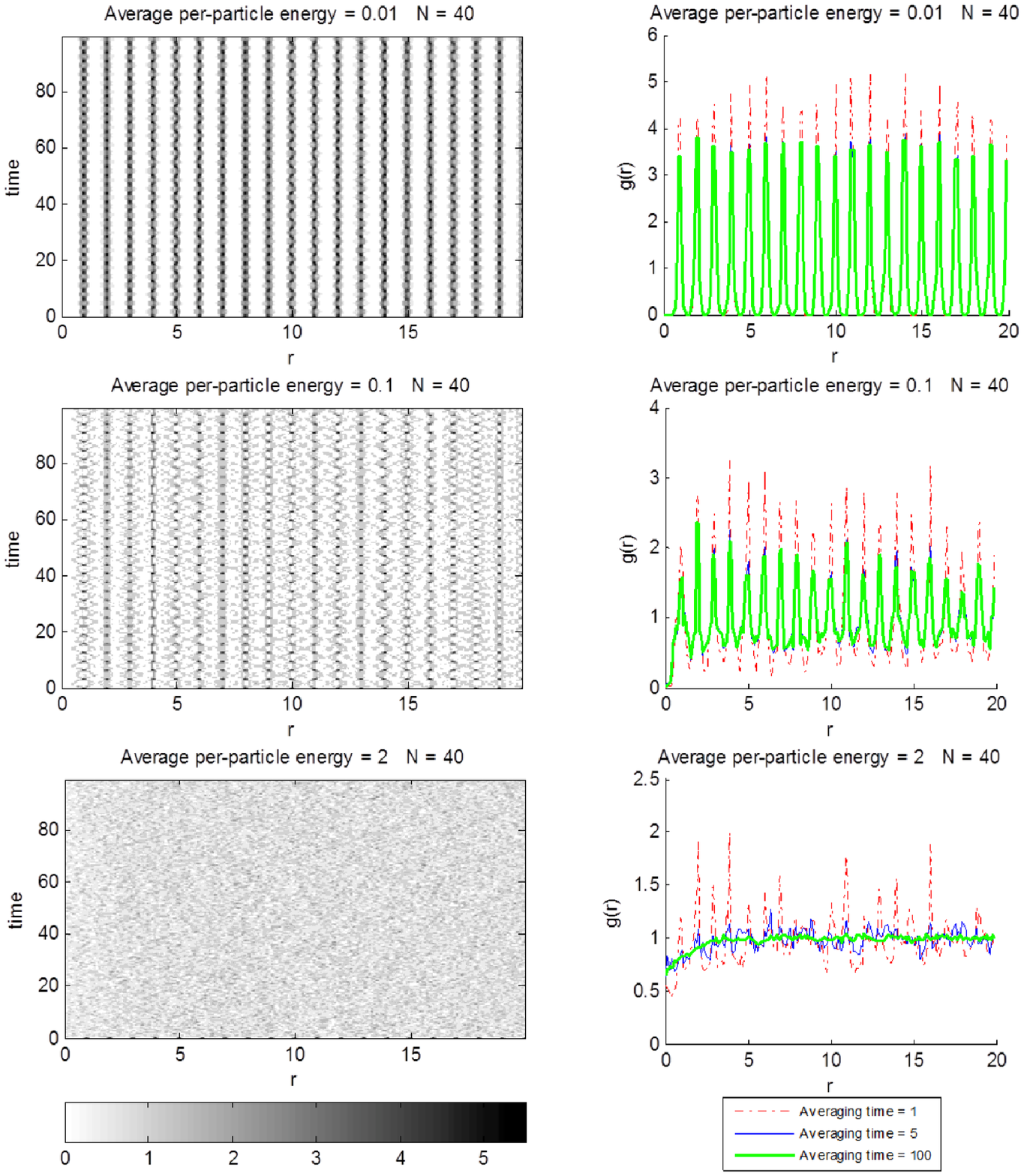}
\caption{\label{fig:CorrelationFunction40} Pair correlation function for the forty-particle system with different per-particle energies. The figures on the left show the time evolution whereas the plots on the right are the time-averaged values for three different averaging times.}
\end{figure*}
\begin{figure}[h]
\includegraphics[scale=0.6]{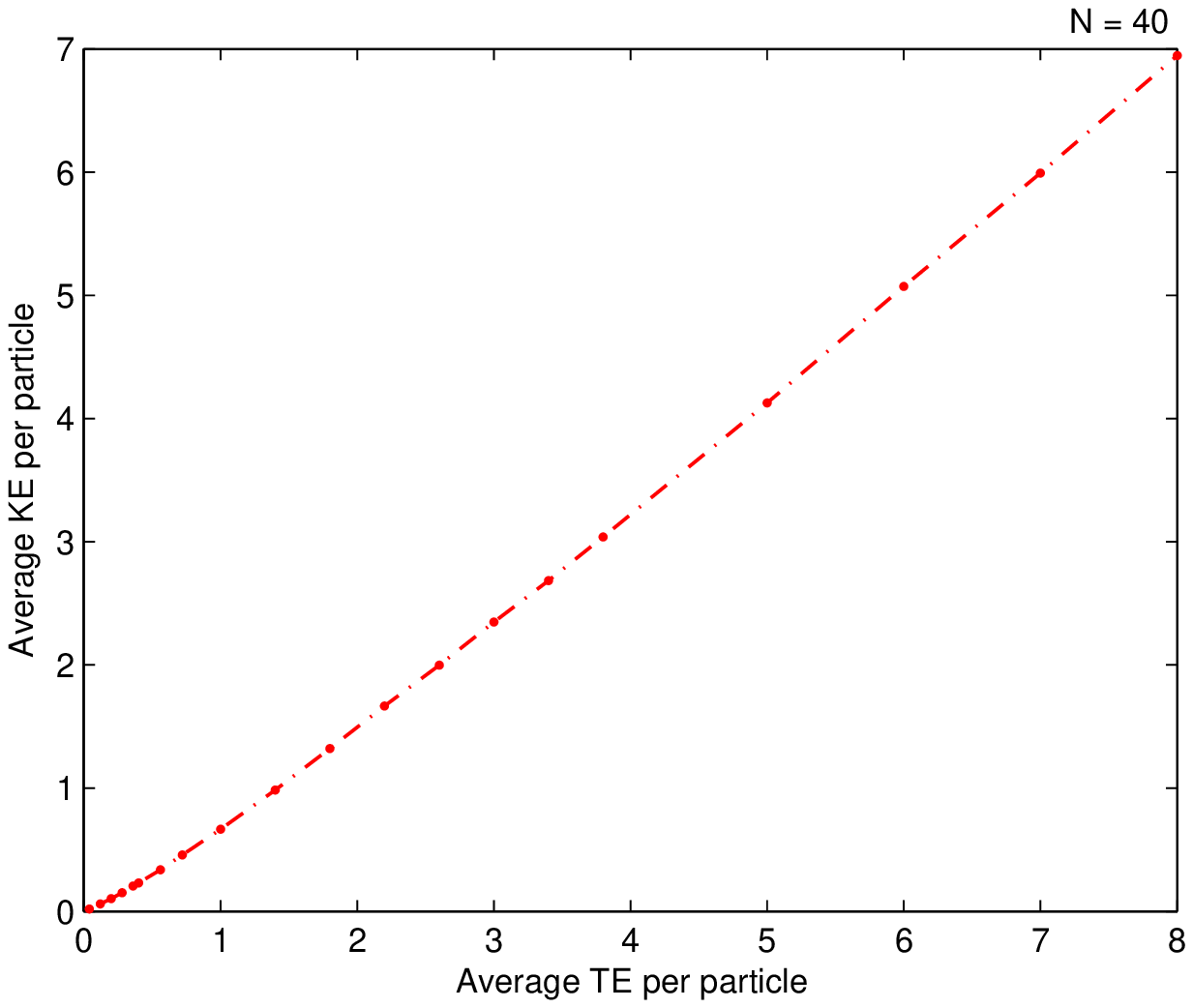}
\caption{\label{fig:KEvsTE40P} Average per-particle kinetic energy plotted against average per-particle total energy for the system of forty particles.}
\end{figure}
\begin{figure}[h]
\includegraphics[scale=0.6]{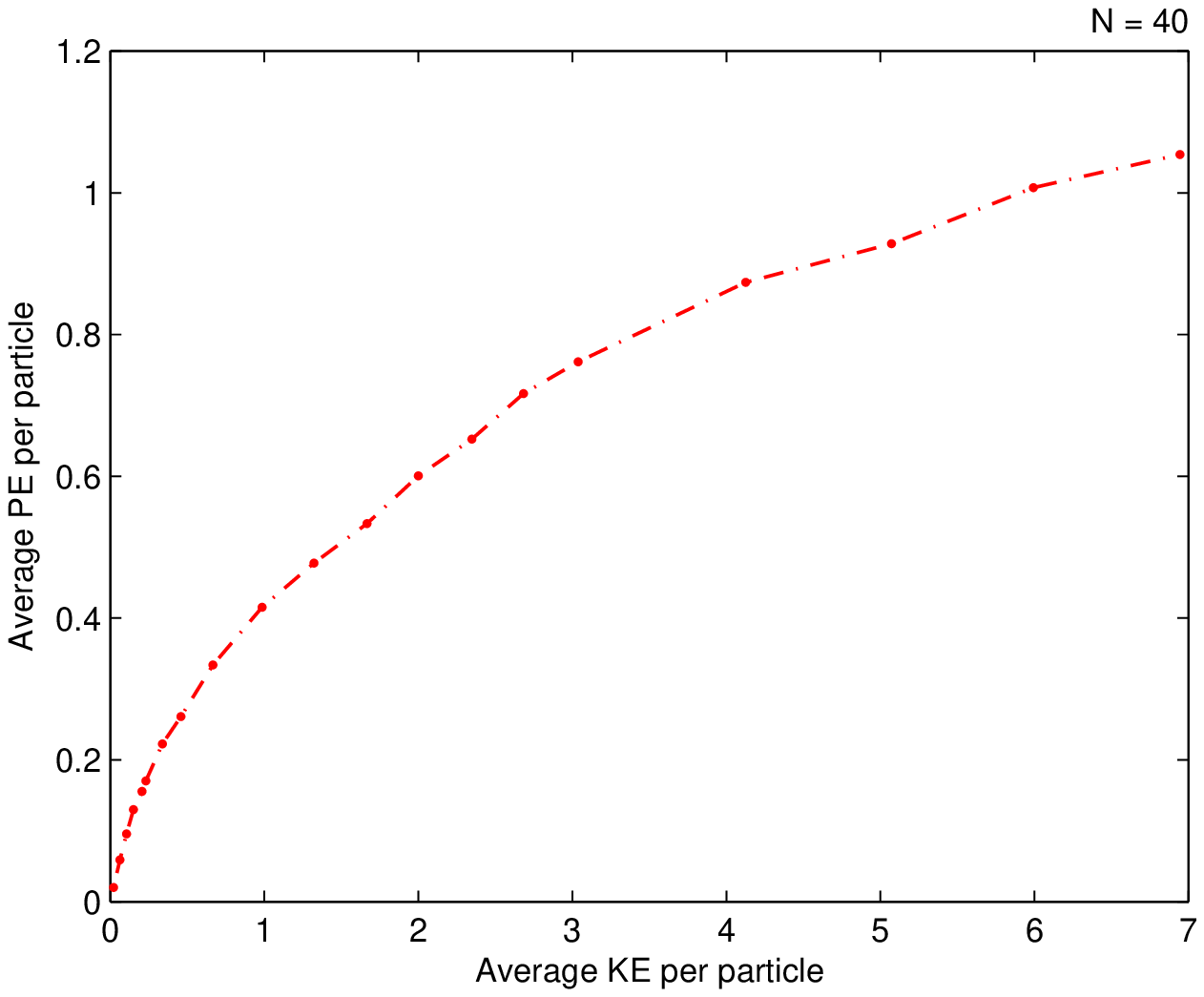}
\caption{\label{fig:PEvsTE40P} Average per-particle potential energy plotted against the average per-particle kinetic energy for the system of forty particles.}
\end{figure}
\begin{figure}[h]
\includegraphics[scale=0.6]{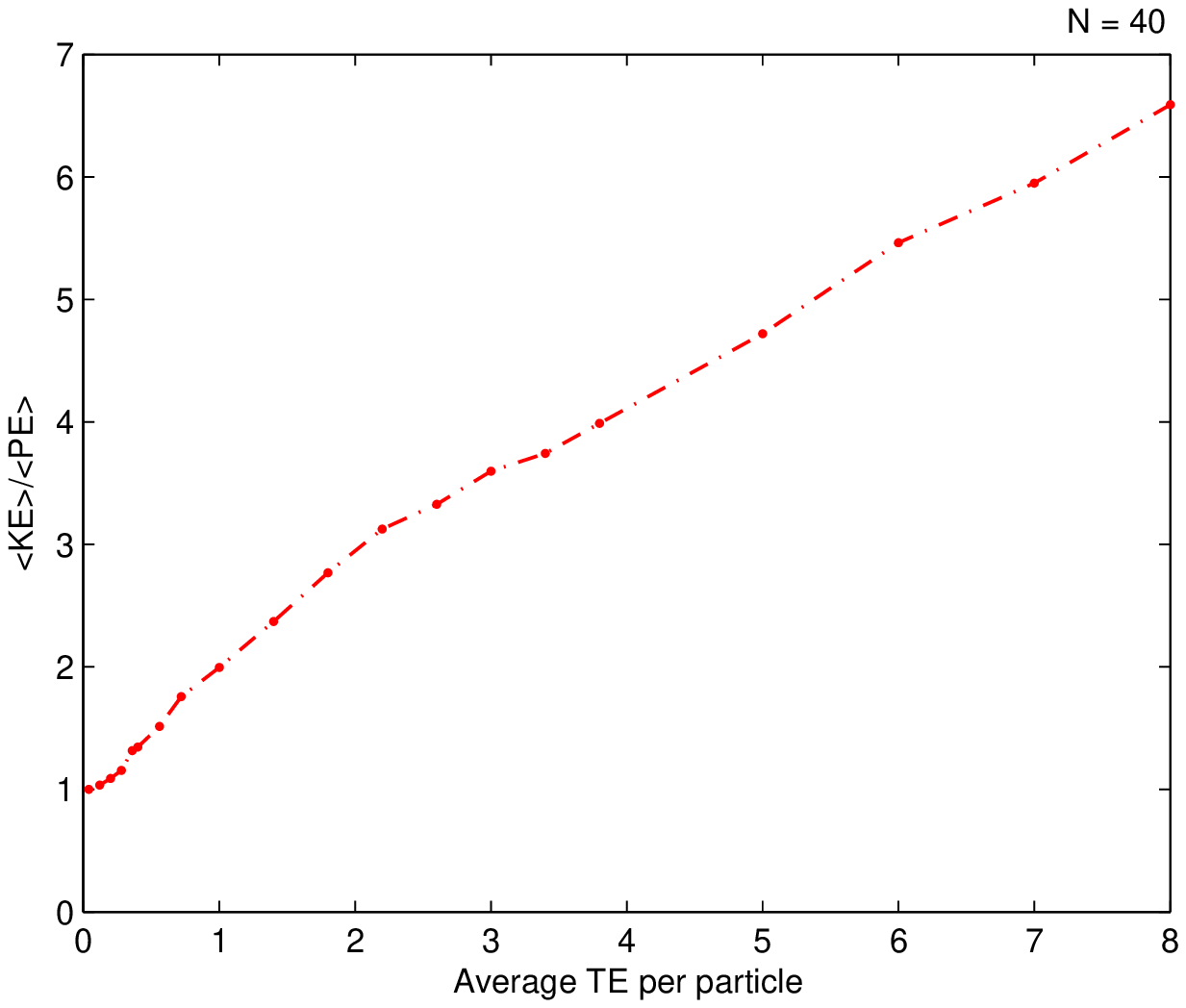}
\caption{\label{fig:VirialVsTE40P} Time averaged virial ratio plotted againt average per-particle energy for the system of forty particles.}
\end{figure}
\begin{figure}[h]
\includegraphics[scale=0.6]{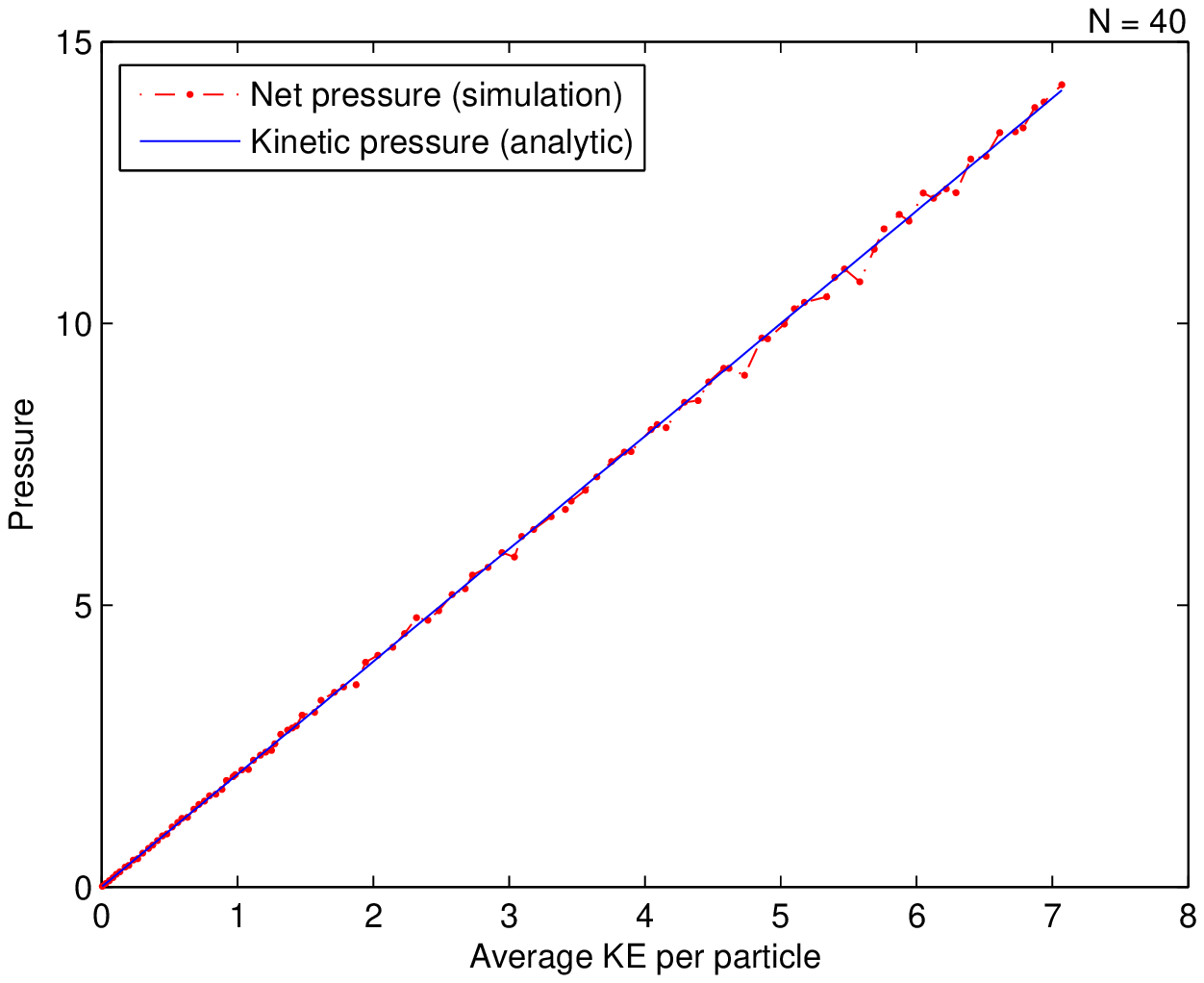}
\caption{\label{fig:PressVsEnergy40P} Average net pressure (from simulation) and kinetic pressure (analytic) plotted against average per-particle kinetic energy for the forty-particle system.}
\end{figure}
\begin{figure}[h]
\includegraphics[scale=0.6]{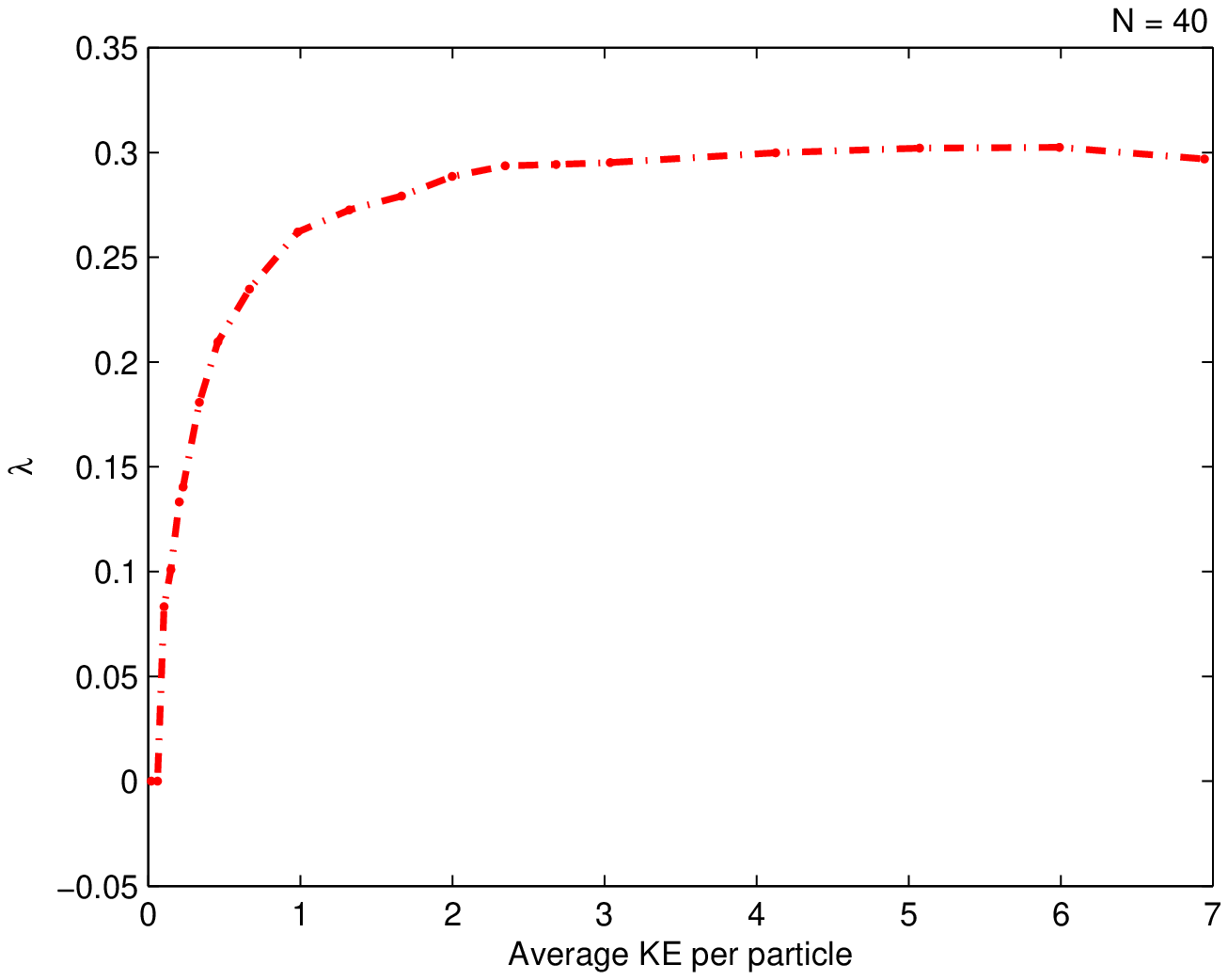}
\caption{\label{fig:LyapVsKE40P} Largest Lyapunov exponent plotted against average per-particle kinetic energy for the forty-particle system.}
\end{figure}
\section{Simulation Results} \label{simulation}
\subsection{Dimensional Parametrization and Initial Conditions}
To elucidate our approach of treating the chaotic dynamics of one-dimensional systems with periodic boundary conditions, we carried out a simulation study of the one-dimensional plasma discussed in Sec. \ref{1DPlasma}. For the purpose of simulation, we parametrized the system in a set of dimensionless units where plasma frequency and the number of particles per unit length are unity. Consequently, the spatial length of the primitive unit cell is numerically equal to the number of particles. The time has been represented in the units of \textquotedblleft per plasma frequency.\textquotedblright The initial conditions are chosen as follows: the particles are placed at the $N$ equilibrium positions obtained from Eq. (\ref{eq_eqmPos}) and are given small random displacements so that the particles do not oscillate in-phase with respect to each other. All the energies are measured with reference to the minimum potential energy for the plasma system. It should be noted that the minimum energy for the system corresponds to the configuration in which the system has no kinetic energy and the distance between any two consecutive particles is the same, that is, the particles are uniformly located in the primitive cell. Velocities are chosen from a random Gaussian distribution and the mean is subtracted from each of the values to obtain zero a center-of-mass velocity. The velocities thus obtained are then scaled such that a desired initial value of per-particle kinetic energy is obtained. For a given set of initial conditions, the time evolution of the particles' positions and velocities are followed using an event-driven algorithm that utilizes the equations derived in Sec. \ref{EqnMotion}. In order to find the largest Lyapunov exponents from Eq. (\ref{eq_LyapExponent}), the phase-space position of the test orbit is determined using Eq. (\ref{eq_testOrbit}) after every iteration.

In statistical mechanics, pressure is an important thermodynamic quantity and several definitions of pressure are available for its evaluation theoretically and in simulation \cite{Choquard1980,Rouet1994}. Taking advantage of our event-driven approach to follow the time evolution, we have devised an algorithm to find the average pressure in the one-dimensional systems with periodic boundary conditions. In our simulation, the pressure is calculated by placing virtual walls uniformly throughout the primitive cell and averaging the momentum transferred from hypothetical elastic collisions to the walls per unit time over a sufficiently long period of time. The wall separation and the averaging time is found by an adaptive algorithm that makes sure that the pressure values have converged to within a pre-assigned small tolerance. The ability to find pressure in simulation is especially advantageous for systems whose exact statistical mechanics have been mathematically challenging.

Another important thermodynamic quantity is the radial distribution function, $g(\textbf{r})$ (also known as the pair correlation function). For a system with bulk density $\rho$, $\rho g(\textbf{r})d\textbf{r}$ is simply the probability of observing a second particle in $dr$ provided that there is a particle at the origin of $\textbf{r}$ \cite{mcquarrie2000}. Our ability to follow the exact time-evolution of the plasma allows us to calculate the radial distribution function. For the one-dimensional periodic plasma, we find the radial distribution function as follows: at the end of the $k$-th iteration (i.e. at time $t=kd\tau$, where $d\tau$ is duration of each iteration) for a particle at $x_j(t)$ on the torus (the primitive cell), the number of particles in a small volume (length) element $\Delta r$ at a distance $r$ on either side of $x_j(t)$ is found. In other words, we look for the number of particles, $\Delta N_j(r,t)$ in a volume of $2\Delta r$ located between ($x_j(t)+r$) and ($x_j(t)+r+\Delta r$) as well as between ($x_j(t)-r-\Delta r$) and ($x_j(t)-r$) on the torus. The radial distribution function is then calculated by averaging for the $N$ particles over a sufficiently long period of time as
\begin{equation} \label{eq_correlationFunct}
g(r) = \lim_{l \to \infty} \frac{\sum_{k=0}^l \sum_{j=1}^N \Delta N_j(r,t=kd\tau)}{(2\Delta r)Nl}
\end{equation}
where $l$ is the number of iterations. Again, the value of $l$ is chosen such that $g(r)$ has converged to a within a small tolerance. For systems with periodic boundary conditions and unit-cell length $2L$, $g(r)$ is periodic with spatial period $L$ and is calculated for $0 \leq r \leq L$.

The event-driven evolution algorithm employed in our simulations allows for efficient determination of the particle-crossing times without performing numerical integration using time steps. However, since we have the analytic solutions of the equations of motion between particle crossings, the entire system can be brought to any desired time without any numerical approximations. Hence, apart from following the evolution of the system crossing-to-crossing, we also find the positions and velocities of the system at fixed intervals of time for the calculation of the largest Lyapunov exponent and the radial distribution function.
 
\subsection{Four-Particle System} \label{simulation_4P}
To examine the properties of the one-dimensional periodic plasma with low number of particles, we carried out a simulation study for a four-particle version of the system. Fig. \ref{fig:Dyn4PLowEnergy} shows the example of time evolution of the four-particle system with low per-particle energies. As we can see, the particles simply oscillate about the equilibrium positions in the primitive cell and no crossing occurs. The time evolution of the four-particle system with high per-particle energy is illustrated in Fig. \ref{fig:Dyn4PHighEnergy}. In the example, the particles not only cross one another, they also cross the boundaries. When a particle crosses one of the primitive-cell boundaries, its replica emerges at the other boundary from the neighboring cell. As mentioned earlier, in order to find unit vector for relative separations in the phase space, the particles also need to be tracked in the extended space not confined to the primitive cell. Fig. \ref{fig:Dyn4PHighEnergyExtSp} shows the evolution of the particles in the extended space for the system depicted in Fig. \ref{fig:Dyn4PHighEnergy}. Evidently, high energy particles will just keep circling on the torus and hence the positions ${{\tilde{x}}_j}(t)$ in the extended space may end up becoming large which could create numerical problems in a computer simulation. However, in a simulation, one does not have to track the particles in the entire extended space; once the phase-space separation has been found for a given iteration and the particles' positions have been deduced in the primitive cell, it suffices to track the particles only in the neighboring cells for the next iteration. Since the unit vector of Eq. (\ref{eq_unitVector}) only depends on the relative separations for a given iteration and since $\tau$ in Eq. (\ref{eq_LyapExponent}) is chosen to be small, the particles in the test system will always be less than $L$ apart with respect to their counterparts in the reference system after the iteration. Fig \ref{fig:CorrelationFunction} shows the evolution of the radial distribution functions $g(r,t)$ (left column) and $g(r)$ averaged for different durations of time (right column) for systems with different energies. It can be seen that with low energy, the particles tend to be located within a small region relative to one another. As the energy is increased, the particles start undergoing crossing and tend to be more evenly spread out in the primitive cell.

Figs. \ref{fig:KEvsTE4P}, \ref{fig:PEvsTE4P} and \ref{fig:VirialVsTE4P} respectively show the average kinetic energy vs. average total energy, average potential energy vs. total energy and virial ratio vs. average total energy on a per-particle basis for the four particle system. Fig. \ref{fig:PressVsEnergy4P} shows the dependence of the average net pressure and the analytic value of the kinetic pressure on the per-particle kinetic energy for the four-particle case. 

An example of the time-evolution of the largest Lyapunov exponent for different values of the perturbation size has been shown for the four-particle system in Fig. \ref{fig:LyapConv4P}. The dependence of the largest Lyapunov on the average per-particle kinetic energy has been depicted in Fig. \ref{fig:LyapVsEnergy4P} for the four-particle system. The largest Lyapunov exponent stays zero for energies that do not allow any interparticle crossings. With an initial increase in energy, more crossings are allowed and the system gets more chaotic. The largest Lyapunov exponent, however, starts dropping back down after reaching a peak for the four-particle system.
 
\subsection{Forty-Particle System} \label{simulation_40P}
To study the dynamics and chaos exhibited by larger plasma systems, we simulated a forty-particle version of the plasma with different initial energies. Fig. \ref{fig:CorrelationFunction40} shows the instantaneous (left column) and the time-averaged (right column) values of the radial distribution function for three different per-particle energies. Like the four-particle system, the particles tend to oscillate about their corresponding equilibrium positions under low-energy conditions and hence the radial distribution function is small for various intervals of $r$. The particles start mixing together as the energy is increased, and eventually, the particles get evenly distributed throughout the unit cell. Figs. \ref{fig:KEvsTE40P}, \ref{fig:PEvsTE40P} and \ref{fig:VirialVsTE40P} respectively show the kinetic energy vs. total energy, potential energy vs. the kinetic energy and the virial ratio vs. total energy where all the energies are measured on a per-particle average basis. Fig. \ref{fig:PressVsEnergy40P} depicts the average net pressure from the simulation as well as the theoretical kinetic pressure as functions of per-particle kinetic energy. Finally, Fig. \ref{fig:LyapVsKE40P} shows the dependence of the largest Lyapunov exponent on the per-particle kinetic energy for the forty-particle system. The behavior of the largest Lyapunov exponent for the forty-particle system is strikingly different from that of the four-particle system depicted in Fig. \ref{fig:LyapVsEnergy4P}; instead of reaching a peak and suddenly dropping back down (as is the case with the four-particle system), the degree of chaos first increases and then is maintained with increasing kinetic energy in the forty-particle system.

\section{Discussion and Conclusions} \label{discussions}
We have presented an appropriate method for representing phase-space vectors that eliminates the possibility of ambiguity or abrupt changes in the phase-space vector components for systems with periodic boundary conditions. By providing the ability to find the largest Lyapunov exponents, the approach provides tools for studying chaos and thermodynamic properties of one-dimensional periodic systems in simulation. Our approach is particularly helpful toward analyzing systems for which analytic treatment of thermodynamics poses challenging mathematical difficulties.

We validated our algorithm with the particular case of a four-particle single component Coulomb system. Moreover, the viability of the method for larger systems has also been demonstrated with a forty-particle system. The study of the four-particle and forty-particle cases provides some interesting insights into the otherwise unexplored chaotic dynamics of such a plasma. Our simulation study indicates that under the low per-particle energy conditions, the particles do not undergo crossings and the ratio of the kinetic energy to potential energy on a per-particle time-average basis is unity, which suggests that the system virializes such that the total energy is shared equally between the kinetic and potential on an average under low-energy conditions. For systems with larger number of particles, kinetic energy tends to dominate the potential energy at a lower per-particle energy indicating that a larger number of particles increases the probability of the onset of interparticle crossings at lower energies. Chaos is not observed in the systems with no crossings and the particles perform pure oscillations about the equilibrium positions.

The radial distribution function in systems with low per-particle energies shows a fluctuating behavior as the radial distance varies. As the energy is increased, the time-averaged radial distribution function, $g(r)$ approaches the behavior of an ergodic system and tends to converge to the expected value of $(N-1)/N$ for large values of the radial distance, $r$. This is evident in both four-particle and forty-particle systems from the plots of the time-averaged $g(r)$ for per-particle energy of $2$ in Figs. \ref{fig:CorrelationFunction} and \ref{fig:CorrelationFunction40}. The behaviors displayed in Figs. \ref{fig:CorrelationFunction} and \ref{fig:CorrelationFunction40} characterize all the randomly sampled versions of the system. However, it is important to note that for special initial conditions that yield periodic or quasi periodic trajectories, this may not be the case. Even though these special cases do not contribute to an ensemble average, the dependence of $g(r)$ on the special initial conditions is worth analyzing and requires further investigation. We plan to study such dependencies in our future work.

Apart from giving an idea about the mixing time and ergodicity in a system, the ability to find the radial distribution function is particularly advantageous from the thermodynamics stand-point; for a system like the one-dimensional plasma whose total potential energy is pair-wise additive, all the thermodynamic parameters can be expressed as a function of $g(r)$ \cite{mcquarrie2000}. Hence, for systems whose exact statistical mechanics continues to be a mathematical challenge, using dynamical simulation to calculate the radial distribution function can play a major role in understanding the  thermodynamics.

Although our study shows that the radial distribution functions follow appreciably similar behavior with changing energy for four-particle and forty-particle systems, the dependencies of their respective largest Lyapunov exponent on the energy are very different (Figs. \ref{fig:LyapVsEnergy4P} and \ref{fig:LyapVsKE40P}). The seemingly sharp peak in the degree of chaos as seen in the four-particle system opens up into a plateau for the forty-particle system. In other words, while the value of the largest Lyapunov exponent starts falling back down with increasing energy after reaching a maximum, the amount of chaos in the forty-particle system is maintained to a high degree after reaching a region of maximum chaos with increasing energy. This change in behavior can be explained basically by looking closely into the four-particle system. As we can see from Fig. \ref{fig:LyapVsEnergy4P}, as the per-particle energy increases, more particles are able to participate in crossings resulting in a diminishing oscillatory behavior and an increasing chaotic behavior. However, once all  four particles have been given sufficient kinetic energy to undergo crossings, a further increase in the kinetic energy will result in the decreased influence of the potential on their motion causing them to start behaving like a free gas. Since the system has periodic boundary conditions, this will mean that the particles are just circling on the torus with the potential having little or negligible effect on the motion. In other words, the four-particle system gradually starts approaching a \textquotedblleft periodic\textquotedblright motion in which the particles circle around the torus in more-or-less fixed amount of time. This behavior slows down the divergence of the test-orbit from its corresponding reference orbit in the phase space and hence the largest Lyapunov exponent starts to decrease. In summation, for our system, a particle with both too much and too little energy results in a decrease in the degree of chaos.

In contrast, in the forty-particle system, once the value of the Lyapunov exponent has reached near the maximum, a further increase in the kinetic energy, on one hand, allows the lower energy particles (which were still trapped in the potential wells of the system) to start undergoing crossings. On the other hand, the particles whose kinetic energies were already high enough to undergo crossing start entering the run-away state where the potential does not have much of an effect on their motion and they start circling the torus. Since there are larger number of particles and there is a richer distribution of velocities, the decrease in chaos because of high energy particles getting even higher energies is compensated by an increase in chaos as a result of particles initially trapped in potential well now being able to undergo crossings. Hence the larger the system, the broader will be the range of kinetic energies for which the degree of chaos will be maintained. Based on this explanation, it can be predicted that a plasma system with initial conditions similar to the ones discussed here will oppose an abrupt change in the value of the largest Lyapunov exponent for all values of energies in the thermodynamic limit ($N \to \infty$) once the Lyapunov exponent has reached near its maximum value.

Due to the lack of an analytic solution for the thermodynamics of a periodic one-component, one-dimensional plasma, the existence of a phase transition cannot be ruled out. Although no credible evidence of a sudden discontinuity in the pressure or its derivative is seen in the four particle case or the forty-particle case, the possibility of such a discontinuity cannot be ignored for the system in the thermodynamic limit and further investigation needs to be done. It should be noted that, in the thermodynamic limit, the behavior of the pressure when plotted against temperature for a system that undergoes a phase transition is expected to show a discontinuity in either the plot itself or in its slope. Hence, for a system with a sufficiently large number of particles, the pressure plots can serve as an indicator of possible phase transitions, and hence the ability to find pressure in simulation adds another tool to examine the thermodynamics in absence of analytical statistical mechanics.

It should be noted that the calculation of the Lyapunov exponents through a geometric analysis of the phase-space trajectories \cite{Caiani1997,Casetti2000} is not applicable in our system. It is evident from Eq. (\ref{eq_netField}) that when a particle crosses another particle, the difference ($N_{left}(x) - N_{right}(x)$) changes abruptly. As a consequence, the first derivative of the velocity exhibits a discontinuity when a particle undergoes a crossing thereby making the tangent space undefined at phase-space positions corresponding to crossings. The non-analyticity in the phase-space trajectories invoked by the discontinuities in the derivatives of the velocities makes the tangent-space approach inapplicable in studying the system. Consequently, we employed the definition of the maximum Lyapunov exponents for flows as expressed in Eq. (\ref{eq_LyapExponent}). In theory, the perturbation, $d_o$ needs to be infinitesimal. However, choosing too small of a value of $d_o$ may result in numerical errors. Our algorithm carefully chooses the value of the perturbation to make sure that the value used is sufficiently small so that the time-evolution of the largest Lyapunov exponent converges to a single behavior. For example, it can be seen from Fig. \ref{fig:LyapConv4P} that the time-evolution behavior and the converged value for the largest Lyapunov exponent do not change considerably by lowering the perturbation size below a sufficiently small value (the value being $1$e$-5$ in the example).

It is evident from Figs. \ref{fig:PressVsEnergy4P} and \ref{fig:PressVsEnergy40P} that the net pressure closely follows the kinetic pressure in both four- and forty-particle versions of the system, even at low energies. However, since the total energy is not proportional to the kinetic energy except at the extremes of large energy, the dependence of pressure on total energy is non-linear, in contrast with a pure gas system. Another point to be be noted is that for the pressure calculation of a purely ergodic system, it would suffice to find the time-average of the momentum transferred to a single wall placed at any location in the primitive cell. However, under low per-particle energy conditions, the system may not undergo interparticle crossings leaving sections of the primitive cell completely unattended. Under such conditions, the average pressure can only be found by placing virtual walls everywhere in the primitive cell.

Finally, it is worth mentioning that no chaos was indicated in a two-particle version of the discussed plasma system for any energy; the possibility of chaotic instability is manifested only in systems with three or more particles. This is consistent with the fact that the dynamics of a conservative one-dimensional system of two particles with no external force can actually be reduced to a one-dimensional, single-particle problem under the constraints of conservation of energy and momentum.

In our ongoing research, we plan to employ the techniques discussed in this paper toward studying the dynamics, stability and indication of possible phase transitions in the one-dimensional periodic versions of single-component and two-component plasma as well as purely gravitational systems.

\begin{acknowledgments}
The authors benefited from helpful conversations with John Hopkins, Igor Prokhorenkov and Jean-Louis Rouet.
\end{acknowledgments}

\bibliography{KumarMiller_ChaoticDynamicsVer1}
\end{document}